\documentclass[aps,prl,floatfix,twocolumn,10pt]{revtex4}

\usepackage{color,amsmath,amssymb,graphicx,latexsym,subfigure}
\usepackage{threeparttable}

\usepackage{titlesec}
\titleformat{\section}[runin]{\normalfont \bfseries}{\thesection}{1em}{}

\newcommand{\blue}[1]{\textcolor{blue}{#1}}

\newcommand{\im}{\text{Im}}

\newcommand{\beq}{\begin{equation}}
\newcommand{\eeq}{\end{equation}}
\newcommand{\bea}{\begin{eqnarray}}
\newcommand{\eea}{\end{eqnarray}}

\newcommand{\gsim}{\lower.7ex\hbox{$\;\stackrel{\textstyle>}{\sim}\;$}}
\newcommand{\lsim}{\lower.7ex\hbox{$\;\stackrel{\textstyle<}{\sim}\;$}}

\newcommand{\be}{\begin{equation}}
\newcommand{\ee}{\end{equation}}
\newcommand{\ba}{\begin{eqnarray}}
\newcommand{\ea}{\end{eqnarray}}

\begin{document}

\title{Probing Axions with Event Horizon Telescope Polarimetric Measurements}

\author{Yifan Chen$^{a,b}$}
\author{Jing Shu$^{a,c,d,e,f}$}
\author{Xiao Xue$^{a,c}$}
\author{Qiang Yuan$^{f,g,h}$}
\author{Yue Zhao$^i$}

\affiliation{
$^a$CAS Key Laboratory of Theoretical Physics, Insitute of Theoretical
Physics, Chinese Academy of Sciences, Beijing 100190, P.R.China\\
$^b$Laboratoire de Physique Th\'eorique et Hautes Energies (LPTHE),\\ UMR 7589,
Sorbonne Universit\'e et CNRS, 4 place Jussieu, 75252 Paris Cedex 05, France\\
$^c$School of Physical Sciences, University of Chinese Academy of
Sciences, Beijing 100049, P.R.China\\
$^d$CAS Center for Excellence in Particle Physics, Beijing 100049, P.R.China\\
$^e$School of Fundamental Physics and Mathematical Sciences, Hangzhou Institute for Advanced Study,\\
University of Chinese Academy of Sciences, Hangzhou 310024, China\\
$^f$Center for High Energy Physics, Peking University, Beijing 100871,
P.R.China\\
$^g$Key Laboratory of Dark Matter and Space Astronomy, Purple Mountain
Observatory, Chinese Academy of Sciences, Nanjing 210008, P.R.China \\
$^h$School of Astronomy and Space Science, University of Science and
Technology of China, Hefei 230026, P.R.China\\
$^i$Department of Physics and Astronomy, University of Utah, Salt Lake
City, UT 84112, USA
}

\begin{abstract}

With high spatial resolution, polarimetric imaging of a supermassive black hole, like M87$^\star$ or Sgr A$^\star$, by the Event Horizon Telescope can be used to probe the existence of ultralight bosonic particles, such as axions. Such particles can accumulate around a rotating black hole through the superradiance mechanism, forming an axion cloud. When linearly polarized photons are emitted from an accretion disk near the horizon, their position angles oscillate due to the birefringent effect when traveling through the axion background. In particular, the observations of supermassive black holes M87$^\star$ (Sgr A$^\star$) can probe the dimensionless axion-photon coupling $c = 2 \pi g_{a \gamma} f_a$ for axions with mass around $O(10^{-20})$~eV ($O( 10^{-17}$)~eV) and decay constant $f_a < O(10^{16})$ GeV, which is complimentary to other axion measurements.

\end{abstract}

\date{\today}

\maketitle
 
\section{Introduction}

The first ever image of the supermassive black hole (SMBH) M87$^\star$ by the Event Horizon Telescope (EHT)~\cite{Akiyama:2019cqa} leads us to a new era of black hole physics. The high spatial resolution makes the direct visual observation of an SMBH and its surroundings possible. While it offers a new way to study the most extreme objects in our universe predicted by Einstein's theory of general relativity, we may wonder what else can we learn, especially for fundamental particle physics, from the rich information extracted from the EHT under such an extreme environment.

The axion is a hypothetical particle beyond the standard model (SM), which was originally motivated by the solution of the strong CP problem~\cite{Peccei:1977hh} in QCD. Beyond the QCD-axion, axion-like particles (ALPs) also generically appear in fundamental theories {\cite{Arvanitaki:2009fg}}, and serve as a viable dark matter candidate \cite{Preskill:1982cy}. There are many search strategies proposed to look for axions, for example, via their conversion into photons \cite{Du:2018uak,Anastassopoulos:2017ftl,Payez:2014xsa,Horns:2012jf}, spectral oscillation or distorsion of photons \cite{Hooper:2007bq, TheFermi-LAT:2016zue,Berg:2016ese}, nuclear magnetic resonance \cite{Graham:2013gfa,Budker:2013hfa}, neutron star mergers \cite{Hook:2017psm} or various table-top experiments~\cite{Stadnik:2013raa,Arvanitaki:2014dfa,Hochberg:2016sqx,TheMADMAXWorkingGroup:2016hpc,Rong:2017wzk,Stadnik:2017hpa,Arvanitaki:2017nhi,Abel:2017rtm,Geraci:2017bmq}.

When the Compton wavelength of an axion is at the same order as the size of a rotating black hole, the axion is expected to develop a large density near the horizon, forming an axion cloud through the superradiance mechanism~\cite{Penrose:1971uk,Press:1972zz,Damour:1976kh,Zouros:1979iw,Detweiler:1980uk,Strafuss:2004qc,Dolan:2007mj,Rosa:2009ei,Dolan:2012yt} (for a review see~\cite{Brito:2015oca}). Such superradiance processes can be tested by black hole spin measurements~\cite{Arvanitaki:2010sy,Arvanitaki:2014wva,Brito:2014wla,Davoudiasl:2019nlo}, gravitational wave signals from bosenova~\cite{Arvanitaki:2010sy,Yoshino:2012kn,Yoshino:2013ofa,Arvanitaki:2014wva,Yoshino:2015nsa,Brito:2017wnc,Brito:2017zvb} or electromagnetic emission from the axion cloud \cite{Rosa:2017ury,Ikeda:2019fvj}. In this letter, we propose a novel way of detecting axion clouds around SMBHs by using the high spatial resolution, polarimetric measurements of the EHT. Our proposed search strategy utilizes the unprecedent capability of the EHT and serves as a complimentary probe.

The recently reported direct image of the shadow of M87$^\star$ illustrates that the EHT is capable of resolving the emission ring close to the event horizon~\cite{Akiyama:2019cqa} where the axion cloud can concentrate. A linearly polarized photon emitted from the innermost region of the accretion disk which lies in a dense axion background experiences the birefringent effect, and its position angle oscillates, with a period being equal to the axion oscillation period~\cite{Carroll:1989vb,Harari:1992ea,Plascencia:2017kca,Ivanov:2018byi,Fujita:2018zaj,Liu:2019brz,Fedderke:2019ajk,Caputo:2019tms}. The amplitude is proportional to the local axion field value. Interestingly, the predicted periodic oscillation of the position angle due to the axion cloud can be the same order as the background Faraday rotation at mm wavelengths for both M87$^\star$ and Sgr A$^\star$~\cite{Macquart:2006zj,Kuo:2014pqa}, making the spatially-resolved polarimetric measurements of the SMBH by the EHT not only optimal to observe the disk magnetic field structure, but also to probe axions.
 
\section{Photon Polarization Variation from Birefringence}
In the presence of an axion-photon interaction and axion potential $V(a)$, the relevant Lagrangian is
\be \mathcal{L}= -\frac{1}{4} F_{\mu\nu} F^{\mu\nu} - \frac{1}{2} g_{a\gamma} a F_{\mu\nu} \tilde{F}^{\mu\nu} + \frac{1}{2} \nabla^\mu a \nabla_\mu a - V(a),\label{Lphoton}\ee
in which $g_{a\gamma}$ is the axion-electromagnetic-field coupling. This modifies the equation of motion for a photon propagating in an axion background field which leads 
to periodic oscillation of a linearly polarized photon's position angle \cite{Carroll:1989vb,Harari:1992ea}. More explicitly, we assume that the variation of the axion 
field in space and time is much slower than the photon's frequency $\omega_\gamma$, i.e., $\mu \ll \omega_\gamma$, where $\mu$ is the axion mass. In later discussions, it will become clear that the spacetime is approximately flat in the region we are interested in. 

In the Lorentz gauge, the modified Maxwell's equation from Eq. (\ref{Lphoton}) for photons propagating along the $z$-axis is
\be \Box A_{\pm} = \pm 2 i g_{a\gamma} [\partial_z a \dot{A}_\pm - \dot{a} \partial_z A_\pm],\ee
where $\pm$ denotes two opposite helicity states with
$A_0 = A_3 = 0, A_\pm = (A_1 \mp i A_2) / \sqrt{2}$, with solutions
\ba A_{\pm} (t, z) &=& A_{\pm} (t', z') \exp\left[-i\omega_\gamma (t - t') 
+ i\omega_\gamma(z - z')\right. \nonumber\\
&\pm& \left. i g_{a\gamma} (a(t, z) - a(t', z'))\right]\blue{.}
\label{photonsolutioninAB} \ea

The leading order effect of the axion background field comes 
from the last term of Eq.~(\ref{photonsolutioninAB}), which results
in a rotation of the position angle for a linearly polarized photon 
\ba \Delta \Theta &=& g_{a\gamma} \Delta a (t_{\rm obs}, \textbf{x}_{\rm obs}; t_{\rm emit}, \textbf{x}_{\rm emit}) \nonumber \\
&=& g_{a\gamma} \int_{\rm emit}^{\rm obs} \ ds\ n^\mu\ \partial_\mu a \nonumber \\
&=& g_{a\gamma} [a(t_{\rm obs}, \textbf{x}_{\rm obs}) - a(t_{\rm emit}, \textbf{x}_{\rm emit})]\blue{.} \label{phaserotation}\ea
Here $n^\mu$ is the null vector along the path. Note that this only depends on the initial and final axion field values. 
 
In the following discussion, we consider photons emitted from the accretion disk near the 
horizon of an SMBH, which are linearly polarized due to synchrotron in the highly ordered magnetic field in the disk \cite{Johnson:2015iwg}, 
and observed at the Earth by, e.g. the EHT. Thus we 
can safely neglect $a (t_{\rm obs}, \textbf{x}_{\rm obs})$ since the axion field can be very large surrounding the SMBH due to superradiance (see below). 
Then Eq. (\ref{phaserotation}) becomes
\bea \Delta \Theta &\simeq& - g_{a\gamma} a(t_{\rm emit}, \textbf{x}_{\rm emit})\nonumber\\ &=& - g_{a\gamma} a_0 (\textbf{x}_{\rm emit}) \cos{ [\omega t_{\rm emit} + \beta(\textbf{x}_{\rm emit})]}.\label{Pangle}\eea
Here $\omega$ is the oscillation frequency of the axion field which depends on the axion mass $\mu$. The amplitude $a_0$ and phase factor $\beta$, are set by the energy density and the phase of the axion cloud at the emission point whose spatial dependence will be discussed later. 

\section{Superradiance and Bosenova}
The axion equation of motion from Eq. (\ref{Lphoton}) in a Kerr background is
\be \Box a = \mu^2 a,\label{KGKerr}\ee
where we take $V(a) = \frac{1}{2} \mu^2 a^2$, and neglect the self-interaction for now. 
After imposing infalling boundary conditions at the black hole horizon, superradiance occurs when the axion frequency $\omega$ is below the critical value
\be |\omega| < \omega_c = \frac{a_J m}{2 M r_+},\label{SRcondition}\ee
where $m$ is the azimuthal number, $a_J$ is the dimensionless spin of the black hole, and $r_+$ is the black hole outer horizon radius. In Planck units ($G_N = c = \hbar =1$), the outer and inner radii are $r_\pm = r_g \left(1\pm \sqrt{1 - a_J^2}\right)$, with $M$ being the black hole mass and $r_g = M$. In this region, one gets a positive value of $\im (\omega)$ which leads to an exponential growth of the wavefunction $a \sim \exp(t/ \tau_{\rm SR})$ with a superradiance timescale $\tau_{\rm SR} \simeq 1/\im (\omega)$.

In Ref.~\cite{Dolan:2007mj}, it was shown that the superradiance takes place efficiently when the Compton wavelength of the axion ($\lambda_C$) is comparable to the size of the rotating black hole within an order one factor: 
\be \frac{r_g}{\lambda_C} = \mu M \equiv \alpha \in (0.1, 1), \ee
where $\alpha \equiv \mu M$ is written in Planck units. In this region, the superradiance rate $\im (\omega)/\mu$ ranges between $10^{-10}$ to $10^{-7}$ according to the simulation in Ref.~\cite{Dolan:2007mj}. For different SMBHs like M87$^{*}$ and Sgr A$^{*}$, the corresponding axion mass windows are different. The axion field produced through superradiance forms a bound state with the SMBH as a ``gravitational atom''. In the $\alpha \ll l$ limit, Eq. (\ref{KGKerr}) for the bound state reduces to the hydrogenic Schr\"odinger equation  with a discrete spectrum
\be 
\label{omega}
\text{Re}(\omega) \simeq \left(1- \frac{\alpha^2}{2\bar{n}^2}\right) \mu,
\ee
where $\bar{n} = n + l +1$ is the principal quantum number. For $m < l$, $\tau_{\rm SR}^{-1}$ is negligible compared to the $m=l$ state and so in the following discussion, we take $m=l$.

So far we have neglected the axion self-interaction, which, as well as the axion mass, arises from instanton corrections induced by associated quantum anomalies. 
The axion potential generically takes the following form, 
\be
\label{classicalaction}
V (a) =  \mu^2 f_a^2 \left(1 - \cos{\frac{a}{f_a}}\right)\blue{,}
\ee
where the leading order expansion around the minimum gives the axion mass term $\frac{1}{2} \mu^2 a^2$.

Since superradiance continually populates the axion cloud, one needs to carefully examine whether self-interaction remains negligible. When $f_a$ is sufficiently large $(> 10^{16}$ GeV)~\cite{Arvanitaki:2014wva}, gravity will always dominate, and the angular momentum of a black hole will decrease until Eq.~(\ref{SRcondition}) no longer holds. Therefore the existence of black holes with high angular momentum has been used to constrain the parameter space of axions in this case~\cite{Arvanitaki:2010sy,Arvanitaki:2014wva,Brito:2014wla,Davoudiasl:2019nlo}. On the other hand, when $f_a$ is small ($<10^{16}$ GeV), the amplitude of the axion field in the axion cloud grows to $f_a$ first before the black hole spin $a_J$ is decreased, and
the self-interactions among these bosons from Eq.~(\ref{classicalaction}) become important compared with gravity, leading to the non-linear regime~\cite{Arvanitaki:2010sy,Arvanitaki:2014wva,Yoshino:2012kn,Yoshino:2013ofa, Yoshino:2015nsa}.

In~\cite{Yoshino:2012kn,Yoshino:2013ofa,Yoshino:2015nsa}, simulations are performed to study this non-linear behavior of the axion cloud. After entering the non-linear regime, the axion cloud either ends as a bosenova explosion or continues to saturate the non-linear region with a steady outflow. In the former case, self-interactions make the axion cloud collapse and the axion field value decreases by an $\mathcal{O}(1)$ factor, then the axion cloud starts to build up again until it reaches the non-linear region at a later time. 
This process could prevent superradiance from persisting, if after a final bosenova explosion a black hole is left unable to reenter the superradiant regime due to environmental effects. In another case, the loss from the gradual scattering towards the far region balances the extraction of the energy from the black hole without triggering an explosion. Although the axion cloud may end up in two very different regimes, the simulations in \cite{Yoshino:2012kn,Yoshino:2015nsa} indicate that the axion field amplitude in the densest region $a_{\rm max}$ is always around $f_a$.  





From Eq.~(\ref{Pangle}), the maximal change of the position angle can be written as:
\be \Delta \Theta_{\rm max} \simeq - b g_{a \gamma} f_a \cos{ [\mu t_{\rm emit} + \beta(|\textbf{x}_{\rm emit}| = r_{\rm max})]}, \ee
by using $a_0 \approx f_a$ and $\omega \approx \mu$ from Eq. (\ref{omega}). Here $b\equiv a_{\rm max}/f_a$, which is an $\mathcal{O}(1)$ number as discussed above.

For general axion, the magnitude of the axion-photon coupling $g_{a \gamma}$ varies depending on the underlying theory and we can redefine $g_{a \gamma}$ to $c_\gamma$ in order to extract the common factors
\bea
g_{a \gamma} \equiv \frac{c}{2 \pi f_a} \equiv \frac{c_\gamma \alpha_{em}} {4 \pi f_a},
\eea 
where $\alpha_{\rm em}$ is the fine-structure constant. In the simplest case, one can imagine $N$ copies of fermions with electric charge $Q$ coupling to the axion with coupling $g_{a f}$, then the axion-photon coupling $g_{a \gamma}$ is induced through the fermion loop and $c_\gamma \sim N Q^2$. In extended theories such as clockwork axions~\cite{Kaplan:2015fuy}, however, the axion-photon coupling $g_{a \gamma}$ could be exponentially large. Consider an $N$-site clockwork model with scalar charge $q$, and let a single set of color neutral vector-like fermions couple at site $M$ ($M < N$).
The axion-photon coupling at low energy is then~\cite{Farina:2016tgd}
\bea
c_\gamma \sim 2 Q^2 q^{N-M}.
\eea
Therefore, we can see that a large $c_{\gamma}$ can compensate the loop suppression factor $\alpha_{em}$ so that $g_{a \gamma}$ can be even larger than $1/f_a$. Notice that our $m_a / f_a$ range is well outside that normally considered for QCD axions, so our axions must be axion like particles (ALPs).



\section{Axion Field Profile}
In this section, we study the axion cloud spatial profile. We focus on the black hole vicinity, especially the region with the ring feature presented by the EHT, i.e. $r_{\rm ring} \simeq 5.5\ r_g$ \cite{Akiyama:2019cqa}. We note that 
the current results published by the EHT group do not have polarization 
information, but the polarization data is expected to be available in the 
future \cite{Doeleman:2019}.


\begin{figure}[htb]
\centering
\includegraphics[width=0.49\textwidth]{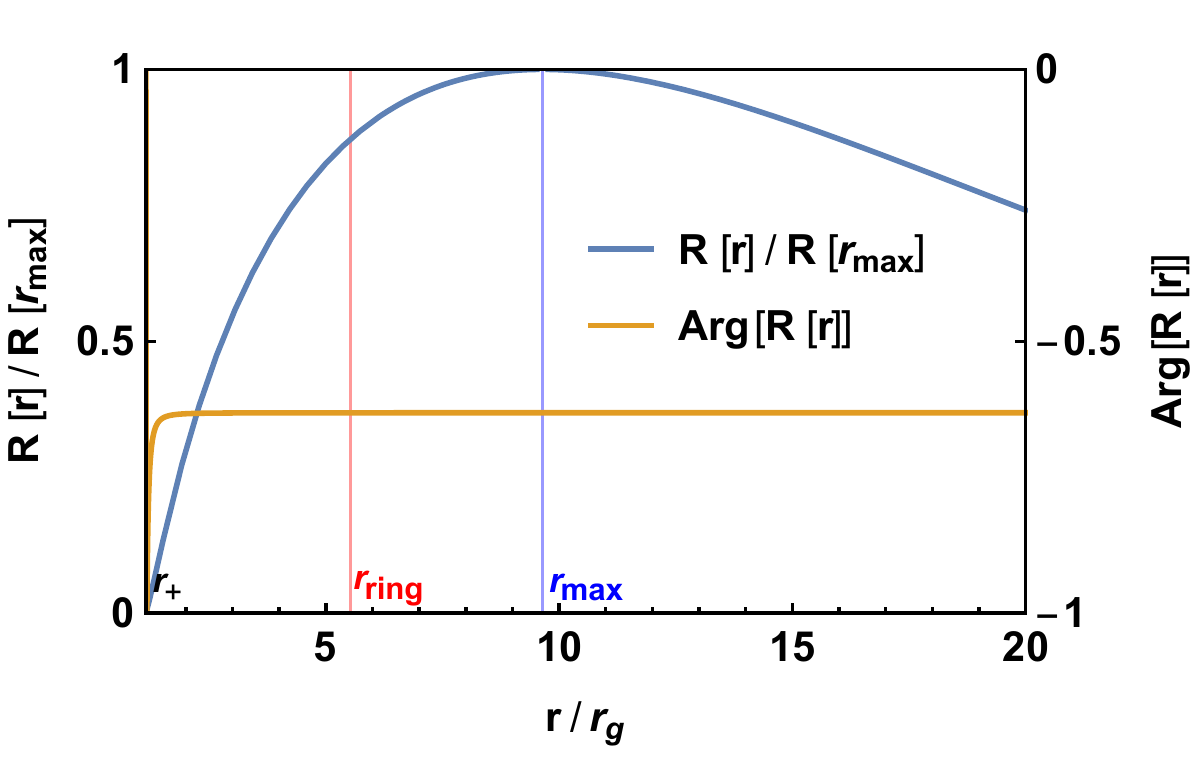}
\caption{The absolute value and the complex phase of $R(r)$ for the $l=1, m=1$ state from Eq. (\ref{Rr}). We take $\alpha$ = 0.4, $a_J = 0.99$.}
\label{figaxionprofile}
\end{figure}

The general solution for Eq. (\ref{KGKerr}) in the Kerr background can be written as
\be a(x^\mu) = e^{-i\omega t} e^{im\phi} S_{lm} (\theta) R_{lm} (r),\label{Kerrsol}\ee
 where $x^\mu = [t, r, \theta, \phi]$ in Boyer-Lindquist coordinates. 
 The $\theta$ dependence is characterized by spheroidal harmonics
$S_{lm} = S^m_l \left(\cos{\theta}, a_J M^2 \sqrt{\omega^2 - \mu^2}\right)$, which simplify to spherical harmonics $Y_l^m$ in the non-rotating or non-relativistic limit.
  
Imposing ingoing boundary condition at the outer horizon $r_+$ and setting the axion field to zero at infinity, the radial part can be written as
 \be
 R(r) = (r - r_+)^{-i\sigma} (r - r_-)^{i\sigma + \chi - 1} e^{qr} \sum_{n=0}^\infty a_n \left(\frac{r-r_+}{r-r_-}\right)^n\blue{,}
 \label{Rr}\ee
 with $\sigma = {2r_+ (\omega - \omega_c)} /{(r_+ - r_-)}$, $q = - \sqrt{\mu^2 - \omega^2}$, and $\chi = {(\mu^2 - 2\omega^2)}/{q}$.
The expansion coefficients, $a_n$, are solved using Leaver's nomenclature in~\cite{Dolan:2007mj}.

One may worry about the non-linear effects discussed in the last section, which come from the generalization of the Klein-Gordon equation to the Sine-Gordon equation arising from Eq. (\ref{classicalaction}). This subtlety is studied in the Appendix of \cite{Yoshino:2012kn} where the deviation from Eq. (\ref{Kerrsol}) is calculated with the Green's function method. It is shown that a perturbative transition to other modes not satisfying the superradiance condition is only significantly induced when there is a bosenova. After the bosenova, these additional modes are expected to fall back into the black hole such that one regains a state similar to the initial perturbative one within a relatively short astrophysical timescale (although this has not yet been unambiguously demonstrated in the simulations~\cite{Yoshino:2012kn,Yoshino:2013ofa,Yoshino:2015nsa}).
Since the superradiance timescale $\tau_{SR}$ is much longer than $\tau_{BN}$, the solution of the Klein-Gordon equation is valid for most of the time during the large period between each bosenova. 


Another issue is whether such a bound state is stable against environmental effects. 
We can estimate the ratio between self and gravitational interactions at a given radius; in the non-relativistic limit, it is maximised around $O(15) r_g$. 
This implies that the point which first triggers bosenova is quite far away from the ring position $r_{\textrm{ring}} \simeq 5.5\ r_g$ observed by EHT, and thus the inner region we have studied should be relatively weakly coupled and stable. 
This back-of-the-envelope estimation is in agreement with simulations (c.f. Fig. 9 of~\cite{Yoshino:2012kn}). 
Therefore, for the state we are likely to be observing long after the final bosenova, 
the axion field value at the ring position should remain of order $f_a$.  


  
In Fig.~\ref{figaxionprofile}, we show the axion field profile for an $l=1, m=1$ state which enjoys the largest superradiance rate. Note that the mixture of different modes with other quantum numbers would not change our following discussing qualitatively since the amplitude for $l=1$ and $m=1$ mode is much higher than other modes and the spatial distributions of the superradiance clouds of different modes are also different. We take $\alpha = 0.4$ and $a_J$ = 0.99 as benchmark. At $r_{\rm ring}= 5.5 r_g$~\cite{Akiyama:2019cqa}, the axion field value is not significantly different from the maximal value, i.e. $R(r_{\rm ring}) \simeq 0.9\ R(r_{\rm max})$. Notice that the complex phase in $R(r)$ is almost a constant for $r>2r_+$ and $S_{lm}$ does not contribute to a complex phase.
Thus the space-dependent complex phase in Eq. (\ref{Pangle}) is 
dominated by $m\phi$ in Eq. (\ref{Kerrsol}),
 \be \beta(\textbf{x}_{\rm emit}) \simeq m\phi.\label{relativephase}
 \ee
The $\Delta \Theta(r,\theta, \phi)$ dependence on the time and spatial position of the radio sources can be obtained from Eq. (\ref{Pangle}), (\ref{Kerrsol}) and (\ref{Rr}),
\bea 
\Delta \Theta (t, r,\theta, \phi) \approx -  \frac{b g_{a\gamma} f_a R_{11} (r)} {R_{11} (r_{\textrm{max}})} \sin \theta \cos{ [\omega t - m \phi]}.
\label{eq:Fig2}
\eea
Taking $b=1$ and $\theta= \pi/2$ which specifies the plane of the accretion disk, the position angle variation is between $\pm 8 c^\circ$. For the time just before the bosenova explosion, it can even reach $\pm 25 c^\circ$. 

\begin{figure}[htb]
\centering
\includegraphics[width=0.48\textwidth]{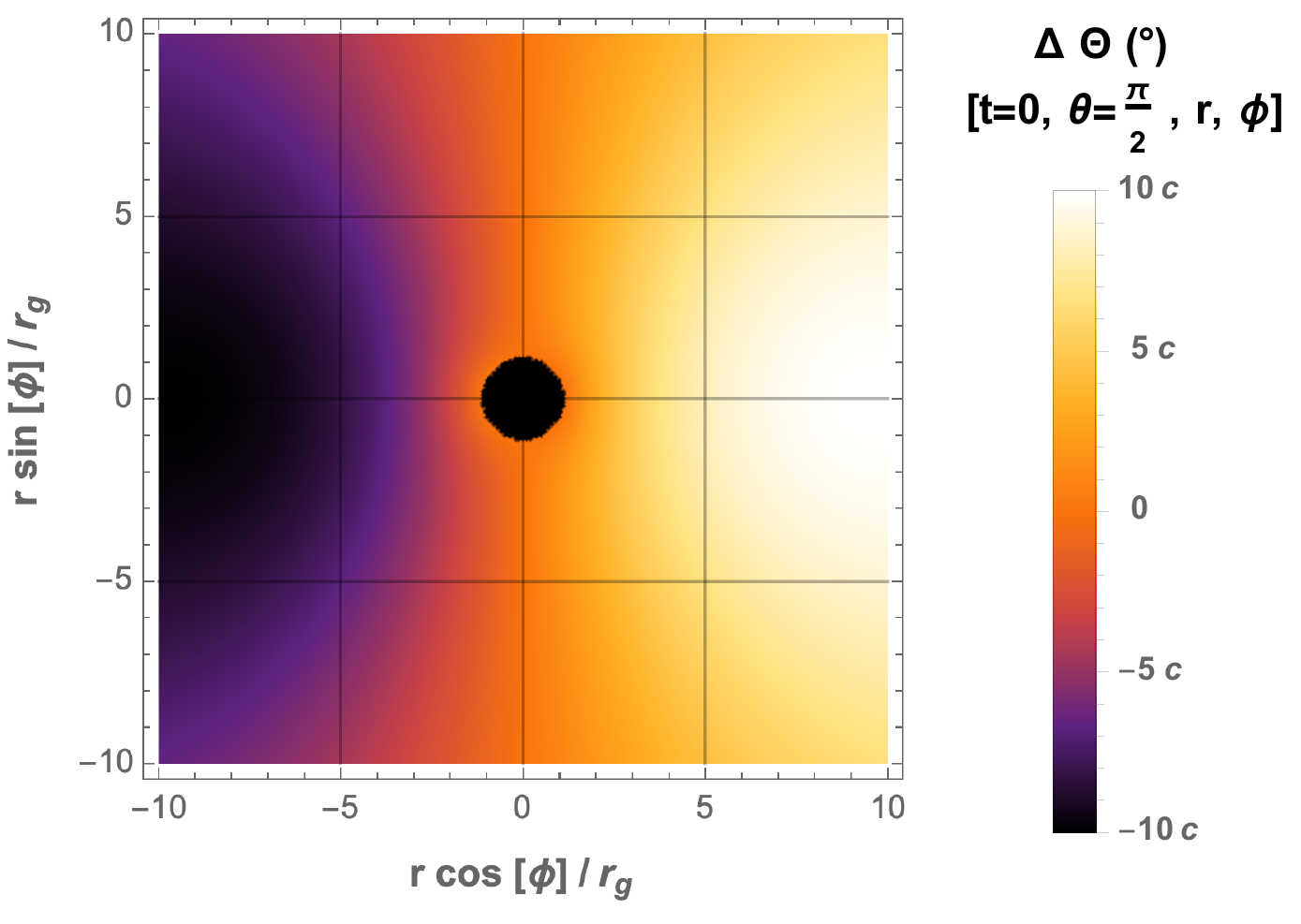}
\caption{$\Delta \Theta (t=0, \theta = \pi/2, r, \phi)$ viewed along the rotating axis of the black hole. The amplitude of oscillation is around $8c^\circ$ 
at $r_{\rm ring}$ for $l=1$, $m=1$, $\alpha = 0.4$, and $a_J = 0.99$.
The region of $r < r_+$ is masked. }
\label{figdeltatheta}
\end{figure}

In Fig.~\ref{figdeltatheta}, assuming that the rotation axis of the disk
points to the observer, we show $\Delta \Theta$ in Eq. (\ref{eq:Fig2}), 
at a fixed time, as a function of position. 
Therefore, both time dependence of the position angle at a fixed spatial point and the position angle as a function of position at a fixed time  
can be used to test the existence of an axion cloud by the high-resolution polarimetric imaging from the EHT.

\section{Detectability}

\begin{table*}[htb]
\begin{center}
\begin{tabular}{ | c | c | c | c | c | c | c | c | c |}		
\hline	
SMBH & M & $a_J$ & $\mu$ range & $\mu$ for $\alpha = 0.4$ & $\tau_a$ & $\tau_{SR}$ \\
\hline
M87$^\star$ & $6.5\times10^9 M_\odot$ & 0.99 & $2.1 \times (10^{-21}\sim10^{-20})$ eV & $8.2\times 10^{-21}$ eV & $5.0\times 10^5$ s & $>1.5\times10^{12}$s  \\
Sgr A$^\star$ &$4.3\times10^6 M_\odot$ & $\cdots$ & $3.1 \times (10^{-18}\sim10^{-17})$ eV & $1.2\times 10^{-17}$ eV & $3.3\times 10^2$ s & $>1.0\times10^{9}$s  \\
\hline
\end{tabular}
\caption{Typical parameters of the axion superradiance of the two SMBHs,
M87$^\star$ and Sgr A$^\star$.}
\label{parameterstable}
\end{center}
\end{table*}

As shown in previous sections, taking $c\sim \mathcal{O}(1)$ as an example, the maximal oscillation amplitude of the position angle is $O(10)^{\circ}$. This is expected to be well within the capability of the EHT. The previous observations of Sgr A$^\star$, with a subset of the EHT configuration and an exposure of tens of minutes, measure the position angle at a precision of $\delta \Theta \sim 3^{\circ}$ \cite{Johnson:2015iwg}. It is reasonable to expect that a better precision can be achieved with even shorter exposure time for the upgraded EHT observations. For Sgr A$^\star$, the expected oscillation period is $100\sim1000$ s (In Table~\ref{parameterstable}, we give a summary of parameters for two SMBHs, M87$^\star$ and Sgr A$^\star$). It might be challenging 
to have a good enough sampling of the observations within one period 
which typically requires an exposure of e.g. tens of seconds. 
The situation for M87$^\star$ is more promising, due to a substantially 
longer oscillation period (a few days). The upcoming analysis of the 
polarization data, particularly for M87$^\star$, should be able to provide valuable information about the possible axion superradiance around the SMBH. 

In order to give a realistic estimation, one needs to take into 
account the spatial resolution of the EHT. The integrals along the $r$ and 
$\theta$ directions always give a positive contribution to the signal, especially when the accretion disk is face-on. Only the average over $\phi$ within the resolution may wash out the axion induced position angle variation. 
The spatial resolution of the M87$^\star$ image is about 20 $\mu$as 
(full width at half maximum; FWHM) \cite{Akiyama:2019cqa}, 
which corresponds to a region of $\sim 2 r_g$ at a distance of $\sim17$ Mpc. 
Assuming a nearly face-on emission disk, which is similar to the 
case of M87$^\star$ with an inclination angle of $\sim 17^\circ$ 
\cite{Akiyama:2019fyp}, this spatial resolution translates to 
$\delta\phi \simeq 4r_g/r_{\rm ring} = 0.7$ rad. Without losing generality, 
taking $\phi = 0$ and considering the average effect within $\delta{\phi}$, 
one obtains the wash-out factor caused by the spatial resolution as
\be \frac{1}{\delta\phi}\int_{-\delta\phi/2}^{\delta\phi/2} \cos ( \mu t + m\phi) d \phi = \frac{\sin{(m\delta\phi/2)}}{m\delta\phi/2} \cos \mu t.
\label{average}
\ee
Without the high spatial resolution provided by the EHT, one has to perform the integration on the whole accretion disk, i.e. $\delta\phi = 2\pi$, 
and the change of the position angle is averaged out. 
For the EHT, taking $m=1$ and $\delta\phi\simeq 0.7$, the wash-out factor is 0.98, and the effect is negligible. 

\begin{figure}[thb]
\centering
\includegraphics[width=0.48\textwidth]{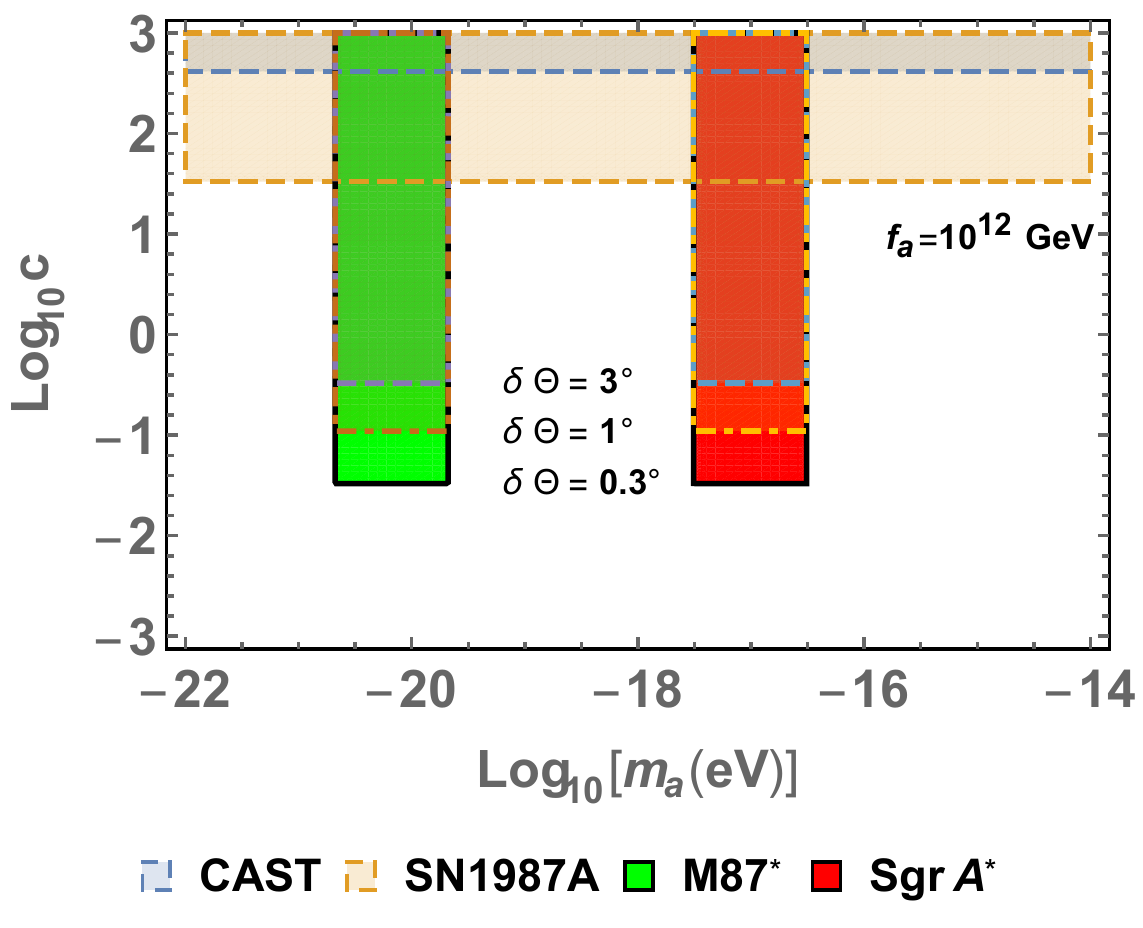}
\caption{The expected parameter space probed by polarimetric observations of M87$^\star$ and Sgr A$^\star$ assuming $b=1$ for position angle precisions 3$^\circ$, 1$^\circ$ and 0.3$^\circ$. We also compare our sensitivity with the bounds from CAST \cite{Anastassopoulos:2017ftl} and Supernova 1987A  \cite{Payez:2014xsa}, assuming $f_a = 10^{12}$ GeV. }
\label{figaxionbound}
\end{figure}


The emission from the accretion disk is expected to be unstable, which
makes the identification of the axion induced signal somehow challenging.
Nevertheless, the astrophysical variability of the disk is usually 
non-periodic. As illustrated in \cite{Johnson:2015iwg}, the position
angle of the linearly polarized emission shows intra-hour variabilities 
in the vicinity of a few Schwarzschild radii. However, the variabilities 
are quite diverse for different observation times. Therefore, the unique 
behavior of periodic oscillation of the position angle due to the axion
field is potentially detectable with high-precision measurements by the 
EHT, through e.g., a periodicity search after Fourier transforming the 
data in the time domain. On the other hand, a null-detection (similar
to the Day 82 observation of SgrA$^\star$~\cite{Johnson:2015iwg} which shows relatively steady position angles~\footnote{Notice that the poor spatial resolution of such an observation makes the time variation of the position angle significantly washed out.}) can give effective 
constraints on the axion parameters.

In Fig.~\ref{figaxionbound}, we show the axion parameter space which is potentially probed by M87$^\star$ and Sgr A$^\star$, assuming $b = 1$ for different position angle precisions. Notice that this method is complementary to the constraints from black hole spin measurements \cite{Arvanitaki:2014wva}, which potentially exclude the region of large $f_a$ ($>10^{16}$ GeV). 
A non-observation of periodic oscillation angles will put an upper bound on the value of $c$ and rule out the corresponding mass window for $f_a < 10^{16}$ GeV.

\section{Conclusion and Discussion}

Dense axion cloud can be induced by rapidly rotating black holes through superradiance. The position angles of linearly polarized photons emitted near the horizon oscillate periodically due to the existence of the axion cloud. A polarimetric measurement with good spatial resolution by e.g., the EHT, is particularly crucial for such a test. The periodic change of the position angle can be tested both temporally and spatially, which would give strong hints of the existence of axion superradiance. 

Our proposed search strategy is complimentary to black hole spin measurements where the axion self-interaction cannot be too strong. 
In addition, when the axion cloud enters the non-linear region, either a drastic bosenova or a steady outflow gives $a_{\rm max}/f_a\sim \mathcal{O}(1)$ for most of the time. Thus our observable does not rely on the detailed dynamics of the axion cloud.

The main constraint on model-dependent factors is the dimensionless axion photon coupling $ c = 2 \pi g_{a \gamma} f_a$, which is rescaled by the scale of the associated symmetry breaking. This makes our experimental constraint unique comparing to other experiments like CAST and SN1987A which only constrain the axion photon coupling $g_{a \gamma}$.


Finally, we note that the position angle oscillation induced by the axion background does not depend on photon frequency. This is a unique property distinct from the Faraday rotation induced by the galactic magnetic field, where the position angle is proportional to the square of photon wavelength. Polarimetric measurements at different frequencies in the future can thus be used to distinguish astrophysical background and improve the sensitivity of tests of the axion superradiance scenario.
\\

\section*{Acknowledgements} 

We are grateful to Nick Houston, Siming Liu, Ru-Sen Lu, and Hirotaka Yoshino for useful discussions. We also thank the anonymous referees for helpful comments and suggestions.
Y.C. is supported by the Labex ``Institut Lagrange de Paris'' (IDEX-0004-02). 
J.S. is supported by the National Natural Science Foundation of China
(NSFC) under Grants No.11947302, No.11690022, No.11851302, No.11675243
and No.11761141011, and by the Strategic Priority Research Program of the 
Chinese Academy of Sciences under Grants No.XDB21010200 and No.XDB23000000. 
Q.Y. is supported by the NSFC under Grants No.11722328, No.11851305, and 
the 100 Talents program of Chinese Academy of Sciences. YZ is supported by U.S. Department of Energy under Award No. DESC0009959.
Y.C. and Y.Z would like to thank the ITP-CAS for their kind hospitality.


\begin{thebibliography}{99}

\bibitem{Akiyama:2019cqa} 
  K.~Akiyama {\it et al.} [EHT Collaboration],
  Astrophys.\ J.\  {\bf 875}, no. 1, L1 (2019).
  K.~Akiyama {\it et al.} [EHT Collaboration],
  Astrophys.\ J.\  {\bf 875}, no. 1, L4 (2019).
  K.~Akiyama {\it et al.} [EHT Collaboration],
  Astrophys.\ J.\  {\bf 875}, no. 1, L6 (2019).


\bibitem{Peccei:1977hh} 
  R.~D.~Peccei and H.~R.~Quinn,
  Phys.\ Rev.\ Lett.\  {\bf 38}, 1440 (1977).
  R.~D.~Peccei and H.~R.~Quinn,
  Phys.\ Rev.\ D {\bf 16}, 1791 (1977).
  S.~Weinberg,
  Phys.\ Rev.\ Lett.\  {\bf 40}, 223 (1978).
  F.~Wilczek,
  Phys.\ Rev.\ Lett.\  {\bf 40}, 279 (1978).

\bibitem{Arvanitaki:2009fg} 
  A.~Arvanitaki, S.~Dimopoulos, S.~Dubovsky, N.~Kaloper and J.~March-Russell,
  Phys.\ Rev.\ D {\bf 81}, 123530 (2010)
  [arXiv:0905.4720 [hep-th]].


\bibitem{Preskill:1982cy} 
  J.~Preskill, M.~B.~Wise and F.~Wilczek,
  Phys.\ Lett.\ B {\bf 120}, 127 (1983).
  L.~F.~Abbott and P.~Sikivie,
  Phys.\ Lett.\ B {\bf 120}, 133 (1983).
  M.~Dine and W.~Fischler,
  Phys.\ Lett.\ B {\bf 120}, 137 (1983).

 \bibitem{Horns:2012jf} 
  D.~Horns, J.~Jaeckel, A.~Lindner, A.~Lobanov, J.~Redondo and A.~Ringwald,
  JCAP {\bf 1304}, 016 (2013)
  [arXiv:1212.2970 [hep-ph]].

\bibitem{Payez:2014xsa} 
  A.~Payez, C.~Evoli, T.~Fischer, M.~Giannotti, A.~Mirizzi and A.~Ringwald,
  JCAP {\bf 1502}, no. 02, 006 (2015)
  [arXiv:1410.3747 [astro-ph.HE]].

\bibitem{Anastassopoulos:2017ftl} 
  V.~Anastassopoulos {\it et al.} [CAST Collaboration],
  Nature Phys.\  {\bf 13}, 584 (2017)
  [arXiv:1705.02290 [hep-ex]].

\bibitem{Du:2018uak}
  N.~Du {\it et al.} [ADMX Collaboration],
  Phys.\ Rev.\ Lett.\  {\bf 120}, no. 15, 151301 (2018)
  [arXiv:1804.05750 [hep-ex]].


\bibitem{Hooper:2007bq} 
  D.~Hooper and P.~D.~Serpico,
  Phys.\ Rev.\ Lett.\  {\bf 99}, 231102 (2007)
  [arXiv:0706.3203 [hep-ph]].

\bibitem{TheFermi-LAT:2016zue} 
  M.~Ajello {\it et al.} [Fermi-LAT Collaboration],
  Phys.\ Rev.\ Lett.\  {\bf 116}, no. 16, 161101 (2016)
  [arXiv:1603.06978 [astro-ph.HE]].

\bibitem{Berg:2016ese} 
  M.~Berg, J.~P.~Conlon, F.~Day, N.~Jennings, S.~Krippendorf, A.~J.~Powell and M.~Rummel,
  Astrophys.\ J.\  {\bf 847}, no. 2, 101 (2017)
  [arXiv:1605.01043 [astro-ph.HE]].

\bibitem{Graham:2013gfa} 
  P.~W.~Graham and S.~Rajendran,
  Phys.\ Rev.\ D {\bf 88}, 035023 (2013)
  [arXiv:1306.6088 [hep-ph]].


\bibitem{Budker:2013hfa} 
  D.~Budker, P.~W.~Graham, M.~Ledbetter, S.~Rajendran and A.~Sushkov,
  Phys.\ Rev.\ X {\bf 4}, no. 2, 021030 (2014)
  [arXiv:1306.6089 [hep-ph]].
    
\bibitem{Hook:2017psm} 
  A.~Hook and J.~Huang,
  JHEP {\bf 1806}, 036 (2018)
  [arXiv:1708.08464 [hep-ph]].

\bibitem{Stadnik:2013raa} 
  Y.~V.~Stadnik and V.~V.~Flambaum,
  Phys.\ Rev.\ D {\bf 89}, no. 4, 043522 (2014)
  [arXiv:1312.6667 [hep-ph]].

  
\bibitem{Arvanitaki:2014dfa} 
  A.~Arvanitaki and A.~A.~Geraci,
  Phys.\ Rev.\ Lett.\  {\bf 113}, no. 16, 161801 (2014)
  [arXiv:1403.1290 [hep-ph]].
  
\bibitem{Hochberg:2016sqx} 
  Y.~Hochberg, T.~Lin and K.~M.~Zurek,
  Phys.\ Rev.\ D {\bf 95}, no. 2, 023013 (2017)
  [arXiv:1608.01994 [hep-ph]].
  
\bibitem{TheMADMAXWorkingGroup:2016hpc} 
  A.~Caldwell {\it et al.} [MADMAX Working Group],
  Phys.\ Rev.\ Lett.\  {\bf 118}, no. 9, 091801 (2017)
  [arXiv:1611.05865 [physics.ins-det]].
  
\bibitem{Rong:2017wzk} 
  X.~Rong {\it et al.},
  Nature Commun.\  {\bf 9}, no. 1, 739 (2018)
  [arXiv:1706.03482 [quant-ph]].
  
\bibitem{Stadnik:2017hpa} 
  Y.~V.~Stadnik, V.~A.~Dzuba and V.~V.~Flambaum,
  Phys.\ Rev.\ Lett.\  {\bf 120}, no. 1, 013202 (2018)
  [arXiv:1708.00486 [physics.atom-ph]].
  
\bibitem{Arvanitaki:2017nhi} 
  A.~Arvanitaki, S.~Dimopoulos and K.~Van Tilburg,
  Phys.\ Rev.\ X {\bf 8}, no. 4, 041001 (2018)
  [arXiv:1709.05354 [hep-ph]].
  
\bibitem{Abel:2017rtm} 
  C.~Abel {\it et al.},
  Phys.\ Rev.\ X {\bf 7}, no. 4, 041034 (2017)
  [arXiv:1708.06367 [hep-ph]].
  
\bibitem{Geraci:2017bmq} 
  A.~A.~Geraci {\it et al.} [ARIADNE Collaboration],
  Springer Proc.\ Phys.\  {\bf 211}, 151 (2018)
  [arXiv:1710.05413 [astro-ph.IM]].
  
\bibitem{Penrose:1971uk} 
  R.~Penrose and R.~M.~Floyd,
  Nature {\bf 229}, 177 (1971).
 
\bibitem{Press:1972zz} 
  W.~H.~Press and S.~A.~Teukolsky,
  Nature {\bf 238}, 211 (1972).
  W.~H.~Press and S.~A.~Teukolsky,
  Astrophys.\ J.\  {\bf 185}, 649 (1973).
  S.~A.~Teukolsky and W.~H.~Press,
  Astrophys.\ J.\  {\bf 193}, 443 (1974).
  
\bibitem{Damour:1976kh} 
  T.~Damour, N.~Deruelle and R.~Ruffini,
  Lett.\ Nuovo Cim.\  {\bf 15}, 257 (1976).
  
\bibitem{Zouros:1979iw} 
  T.~J.~M.~Zouros and D.~M.~Eardley,
  Annals Phys.\  {\bf 118}, 139 (1979).
  
\bibitem{Detweiler:1980uk}
  S.~L.~Detweiler,
  Phys.\ Rev.\ D {\bf 22} (1980) 2323.
  
\bibitem{Strafuss:2004qc} 
  M.~J.~Strafuss and G.~Khanna,
  Phys.\ Rev.\ D {\bf 71}, 024034 (2005)
  [gr-qc/0412023].
  
\bibitem{Dolan:2007mj} 
  S.~R.~Dolan,
  Phys.\ Rev.\ D {\bf 76}, 084001 (2007)
  [arXiv:0705.2880 [gr-qc]].
  
\bibitem{Rosa:2009ei} 
  J.~G.~Rosa,
  JHEP {\bf 1006}, 015 (2010)
  [arXiv:0912.1780 [hep-th]].
  
\bibitem{Dolan:2012yt} 
  S.~R.~Dolan,
  Phys.\ Rev.\ D {\bf 87}, no. 12, 124026 (2013)
  [arXiv:1212.1477 [gr-qc]].
  
\bibitem{Brito:2015oca} 
  R.~Brito, V.~Cardoso and P.~Pani,
  Lect.\ Notes Phys.\  {\bf 906}, pp.1 (2015)
  [arXiv:1501.06570 [gr-qc]].
  

 
\bibitem{Arvanitaki:2010sy} 
  A.~Arvanitaki and S.~Dubovsky,
  Phys.\ Rev.\ D {\bf 83}, 044026 (2011)
  [arXiv:1004.3558 [hep-th]].
 
\bibitem{Arvanitaki:2014wva} 
  A.~Arvanitaki, M.~Baryakhtar and X.~Huang,
  Phys.\ Rev.\ D {\bf 91}, no. 8, 084011 (2015)
  [arXiv:1411.2263 [hep-ph]].
  
\bibitem{Brito:2014wla} 
  R.~Brito, V.~Cardoso and P.~Pani,
  Class.\ Quant.\ Grav.\  {\bf 32}, no. 13, 134001 (2015)
  [arXiv:1411.0686 [gr-qc]].

\bibitem{Davoudiasl:2019nlo} 
  H.~Davoudiasl and P.~B.~Denton,
  Phys.\ Rev.\ Lett.\  {\bf 123}, no. 2, 021102 (2019)
  [arXiv:1904.09242 [astro-ph.CO]].
  

\bibitem{Yoshino:2012kn} 
  H.~Yoshino and H.~Kodama,
  Prog.\ Theor.\ Phys.\  {\bf 128}, 153 (2012)
  [arXiv:1203.5070 [gr-qc]].

\bibitem{Yoshino:2013ofa} 
  H.~Yoshino and H.~Kodama,
  PTEP {\bf 2014}, 043E02 (2014)
  [arXiv:1312.2326 [gr-qc]].

\bibitem{Yoshino:2015nsa} 
  H.~Yoshino and H.~Kodama,
  Class.\ Quant.\ Grav.\  {\bf 32}, no. 21, 214001 (2015)
  [arXiv:1505.00714 [gr-qc]].



  
\bibitem{Brito:2017wnc} 
  R.~Brito, S.~Ghosh, E.~Barausse, E.~Berti, V.~Cardoso, I.~Dvorkin, A.~Klein and P.~Pani,
  Phys.\ Rev.\ Lett.\  {\bf 119}, no. 13, 131101 (2017)
  [arXiv:1706.05097 [gr-qc]].
  
\bibitem{Brito:2017zvb} 
  R.~Brito, S.~Ghosh, E.~Barausse, E.~Berti, V.~Cardoso, I.~Dvorkin, A.~Klein and P.~Pani,
  Phys.\ Rev.\ D {\bf 96}, no. 6, 064050 (2017)
  [arXiv:1706.06311 [gr-qc]].
    
    
\bibitem{Rosa:2017ury} 
  J.~G.~Rosa and T.~W.~Kephart,
  Phys.\ Rev.\ Lett.\  {\bf 120}, no. 23, 231102 (2018)
  [arXiv:1709.06581 [gr-qc]].
    
\bibitem{Ikeda:2019fvj} 
  T.~Ikeda, R.~Brito and V.~Cardoso,
  Phys.\ Rev.\ Lett.\  {\bf 122}, no. 8, 081101 (2019)
  [arXiv:1811.04950 [gr-qc]].
  
  

\bibitem{Carroll:1989vb}
  S.~M.~Carroll, G.~B.~Field and R.~Jackiw,
  Phys.\ Rev.\ D {\bf 41}, 1231 (1990).
  S.~M.~Carroll and G.~B.~Field,
  Phys.\ Rev.\ D {\bf 43}, 3789 (1991).

\bibitem{Harari:1992ea} 
  D.~Harari and P.~Sikivie,
  Phys.\ Lett.\ B {\bf 289}, 67 (1992).

\bibitem{Plascencia:2017kca} 
  A.~D.~Plascencia and A.~Urbano,
  JCAP {\bf 1804}, no. 04, 059 (2018)
  [arXiv:1711.08298 [gr-qc]].
  
\bibitem{Ivanov:2018byi} 
  M.~M.~Ivanov, Y.~Y.~Kovalev, M.~L.~Lister, A.~G.~Panin, A.~B.~Pushkarev, T.~Savolainen and S.~V.~Troitsky,
  JCAP {\bf 1902}, no. 02, 059 (2019)
  [arXiv:1811.10997 [astro-ph.CO]].
  




\bibitem{Fujita:2018zaj} 
  T.~Fujita, R.~Tazaki and K.~Toma,
  Phys.\ Rev.\ Lett.\  {\bf 122}, no. 19, 191101 (2019)
  [arXiv:1811.03525 [astro-ph.CO]].
  
\bibitem{Liu:2019brz} 
  T.~Liu, G.~Smoot and Y.~Zhao,
  arXiv:1901.10981 [astro-ph.CO].
  
  
\bibitem{Fedderke:2019ajk} 
  M.~A.~Fedderke, P.~W.~Graham and S.~Rajendran,
  Phys.\ Rev.\ D {\bf 100}, no. 1, 015040 (2019)
  [arXiv:1903.02666 [astro-ph.CO]].
  

\bibitem{Caputo:2019tms} 
  A.~Caputo, L.~Sberna, M.~Frias, D.~Blas, P.~Pani, L.~Shao and W.~Yan,
  arXiv:1902.02695 [astro-ph.CO].
  
\bibitem{Macquart:2006zj} 
  J.~P.~Macquart, G.~C.~Bower, M.~C.~H.~Wright, D.~C.~Backer and H.~Falcke,
  Astrophys.\ J.\  {\bf 646}, L111 (2006)
  [astro-ph/0606381].

\bibitem{Kuo:2014pqa} 
  C.~Y.~Kuo {\it et al.},
  Astrophys.\ J.\  {\bf 783}, L33 (2014)
  [arXiv:1402.5238 [astro-ph.GA]]  


\bibitem{Johnson:2015iwg} 
  M.~D.~Johnson {\it et al.},
  Science {\bf 350}, no. 6265, 1242 (2015)
  [arXiv:1512.01220 [astro-ph.HE]].

\bibitem{Kaplan:2015fuy} 
  D.~E.~Kaplan and R.~Rattazzi,
  Phys.\ Rev.\ D {\bf 93}, no. 8, 085007 (2016)
  [arXiv:1511.01827 [hep-ph]].
    
\bibitem{Farina:2016tgd} 
  M.~Farina, D.~Pappadopulo, F.~Rompineve and A.~Tesi,
  JHEP {\bf 1701}, 095 (2017)
  [arXiv:1611.09855 [hep-ph]].
    
  \bibitem{Doeleman:2019}
S.~Doeleman,
\url{https://iopscience.iop.org/journal/2041-8205/page/Focus_on_EHT}
 
  
\bibitem{Akiyama:2019fyp} 
  K.~Akiyama {\it et al.} [EHT Collaboration],
  Astrophys.\ J.\  {\bf 875}, no. 1, L5 (2019).
  

  
\end{thebibliography}
\end{document}